\documentclass[prd,aps,a4paper,floatfix]{revtex4}

\usepackage{amsmath}
\usepackage{graphicx}
\usepackage{dcolumn}
\usepackage{bm}

\def\mnras{Mon. Not. Roy. Astron. Soc.}

\def\prd{Phys. Rev. {\bf D}}
\def\physrep{Physics Reports }
\def\procspie{Proceedings of the SPIE}
\def\apj{Astrophys. Journal}
\def\apjs{Astrophys. Journal Supp.} 
\begin{document}


\title{The large scale CMB cut-off and the tensor-to-scalar ratio}

\author{Gavin Nicholson}
\author{Carlo R. Contaldi}%
\affiliation{%
Theoretical Physics, Blackett Laboratory, Imperial College, Prince
 Consort Road, London, SW7 2BZ, U.K.}%

\date{\today}

\begin{abstract}
  We show that if inflation lasted just longer than the required 60 or
  so e-folds $N$ both scalar and tensor contributions to the Cosmic
  Microwave Background (CMB) power spectra are expected to show a
  cut-off. However the behaviour of the scalar-to-tensor ratio on
  large scale depends on whether inflation is preceded by a Kinetic
  Dominated (KD) or Radiation Dominated (RD) stage. Future experiments
  may be able to distinguish between the two behaviours and thus shed
  light on the nature of the low-$\ell$ cutoff. In particular if the
  cut-off is due to a KD stage the ratio is expected to grow on large
  scales. If observed this would challenge our current understanding
  of the overall probability of inflation lasting for $N$ greater than
  60.
\end{abstract}

\pacs{Valid PACS appear here}
\maketitle

\section{Introduction}\label{sec:intro}

The anomalously low quadrupole of the angular spectrum of the CMB
\cite{Hinshaw06} has generated interest in models of inflation which
can accommodate a cut-off in the spectrum of curvature
perturbations. Rather than discarding the inflationary paradigm
altogether these models aim to incorporate large scale departures from
scale invariance by minimal modifications of the scenarios. After all,
inflation has been very successful in predicting a nearly scale
invariant, Gaussian and coherent spectrum of primordial curvature
perturbations which seems to fit all other observable scales extremely
well. In addition the statistical significance of the cut-off itself
is questionable and has been much discussed in the literature (see
e.g. \cite{efstathiou,costa,magueijo,bielewicz}) and questions remain
as to whether the observations are affected by foregrounds on these
scales \cite{bridle,schwarz,slosar,prunet}.

A common scenario used to obtain a cut-off is to assume that inflation
lasted just long enough ($\sim 60$ $e$-foldings)
\cite{Contaldi03,Cline03,Kaloper04,Piao04,Kinney06}. The inflating
stage is either preceded by kinetic dominated (KD) or ``fast-rolling''
stage or a radiation dominated (RD) epoch. The time at which the
transition between the pre-inflating and inflating epoch occurs is
tuned such that its imprint is observed in the modes that are entering
the horizon at the present time i.e. the largest observable
scales. Since these modes exited the horizon when slow roll was not
established their evolution is more complicated than in the
standard approximation but a generic feature in many such models is a
suppression of the scalar power in modes larger than the cut-off
scale.

The detailed nature of the cut-off in these models depends on the
initial conditions assumed for both the background and the inflaton
perturbations themselves. The initial conditions are not defined
as well as in scenarios where inflation lasts for much longer than $N\sim
60$. In this case there is no small-scale, adiabatic limit for the
modes of interest. Thus some assumptions have to be made about their
initial normalisation.

Another drawback of these models is that the reduced number of
$e$-foldings involved may not be sufficient to solve the horizon and
flatness problems. Thus some residual fine-tuning is required under
these conditions. A possible solution to this could be to assume
that a previous stage of inflation occurred with an intervening
non-inflating stage between the end of the first stage and the last
bout of inflation. This last stage is the one that resulted in portion
of the spectrum that is observable today.

By far the biggest tuning problem in the models is related to
the vanishing probability of inflationary solutions with such limited
number of $e$-folding. This has been the subject of a long debate. In
the context of chaotic inflation models \cite{chaotic} the probability
of having an inflationary stage that lasts {\sl just} long enough is
low. Most solutions would lead to much longer periods of inflation
\cite{Kofman02}. According to this view the cut-off models face a
severe fine-tuning problem. However other approaches based on the
canonical measure of inflation \cite{Gibbons86,Hawking88} seems to lead to a
conflicting picture, with the probability of a long inflationary stage
exponentially suppressed by a factor $e^{-3N}$ in \cite{Gibbons06}.

The landscape paradigm of string theory vacua which has emerged in
recent years may also imply that the number of inflationary
$e$-foldings is suppressed. In particular \cite{freivogel} have
suggested that inflation proceeds after the inflaton tunnels
\cite{coleman} from one of the many metastable vacua in the landscape
to a broad one where the inflaton is initially kinetic dominated and
then slow-rolls as in the conventional picture. The interesting aspect
of this scenario is that statistical arguments of landscape vacua
\cite{douglas} result in a highly suppressed probability of long
lasting inflation. Thus, if confirmed, the cut-off will certainly fuel
the debate on the overall measure of inflation but may also reaffirm
our current understanding of the low energy picture of string theory.

In this work we show how the polarisation signature of the CMB may
lead to further characterisation of the cut-off. This occurs in two
ways. Firstly, the tensor modes themselves should show signs of out of
slow-roll evolution on the same scales as the scalar. This may shed
more light on the statistical or physical nature of the observed
anomaly. Secondly, the behaviour of the cut-off and in particular the
 tensor-to-scalar ratio on large scales depends on the exact scenario
preceding the slow-roll phase. The tensor contribution to the CMB is
directly observable in the B-mode, or curl-like,
polarisation. Experiments are already targeting this signature with
many more to come and so it is timely to examine whether the
observations will shed more light on the nature of the inflaton.

The {\sl paper} is organised as follows. We start in
section~\ref{sec:CMB} by showing how the tensor-to-scalar ratio
differs if the pre-inflationary stage is KD or RD. In
section~\ref{sec:observe} we show that the difference in behaviour may
be within observable reach of future all-sky polarisation sensitive
experiments. In section~\ref{sec:disc} we end with a discussion of the
implication of a confirmation of the cut-off for inflationary measures
and the string theory landscape.

\section{Short inflation and the CMB cut-off}\label{sec:CMB}

We examine the evolution of perturbations in short inflation scenarios
by numerical evolution from two separate initial conditions. The first
is a ``fast-to-slow-roll'' scenario in which the inflaton starts in a
kinetic energy dominated phase and transitions into a slow--roll
trajectory. This type of scenario was introduced in  \cite{Contaldi03}
as a possible origin of the low-$\ell$ anomaly. The second scenario is
one where the evolution starts in a radiation dominated phase and
transitions to slow--roll inflation.

The first case encompasses models where the observable universe
started inflating with stochastic initial conditions in the inflaton
field phase space. The simplest example of this is the  chaotic inflation model
\cite{chaotic} with a quadratic potential. The
second scenario is more suitable for models with a radiation
dominated epoch before the inflationary phase responsible for
solving the cosmological problems. This would occur, for example, in
models where a number of inflationary stages occur each separated by a
reheating stage and further radiation domination. In both cases we
evolve the perturbations in a flat, isotropic background for
simplicity \footnote{see e.g. \cite{emir} for more general initial
  conditions.}.  

\subsection{Kinetic dominated initial conditions}

The effect of fast rolling initial conditions on the scalar mode power
spectrum was described in \cite{Contaldi03}. We will revisit the
scenario in this work in order to derive the expected signature of the
tensor modes. We chose two simple models with inflationary potentials
$V(\phi) = m^2\phi^2/2$ and $V(\phi) = \lambda \phi^4/4$.

We calculate the power spectrum of primordial, super-horizon modes at
a number of wavelengths by numerically evolving the system of
background and perturbation equations. We denote scalar perturbations
of the FRW metric in the conformal Newtonian gauge as
\begin{equation}
  ds^2 = a^2(\eta)\left[(1+2\Phi)d\eta^2 - (1-2\Phi)\delta_{ij}dx^idx^j\right],
\end{equation}
where $\eta$ is conformal time, $\Phi$ is the curvature perturbation
and we have assumed zero anisotropic shear in the einstein equations
such that we can equate the curvature and potential
perturbations. Tensor perturbations to the metric are denoted in terms
of the traceless, transverse tensor $h_{ij}$ (with $i,j$ spanning
spatial indexes only)
\begin{equation}
  ds^2 =  a^2(\eta)\left[d\eta^2 - (\delta_{ij}-h_{ij})dx^idx^j\right].
\end{equation}

We recast the scalar perturbation variable into the gauge invariant Mukhanov
variable \cite{Mukhanov} 
\begin{equation}
  u = a\left( \delta \phi + \frac{\phi'}{\cal H}\Phi\right),
\end{equation} 
where primes denote derivatives with respect to conformal time, $\delta\phi$
is the inflaton perturbation and ${\cal H} = a'/a$.

\begin{figure}[t]
\centering
\includegraphics[width=9cm,angle=270]{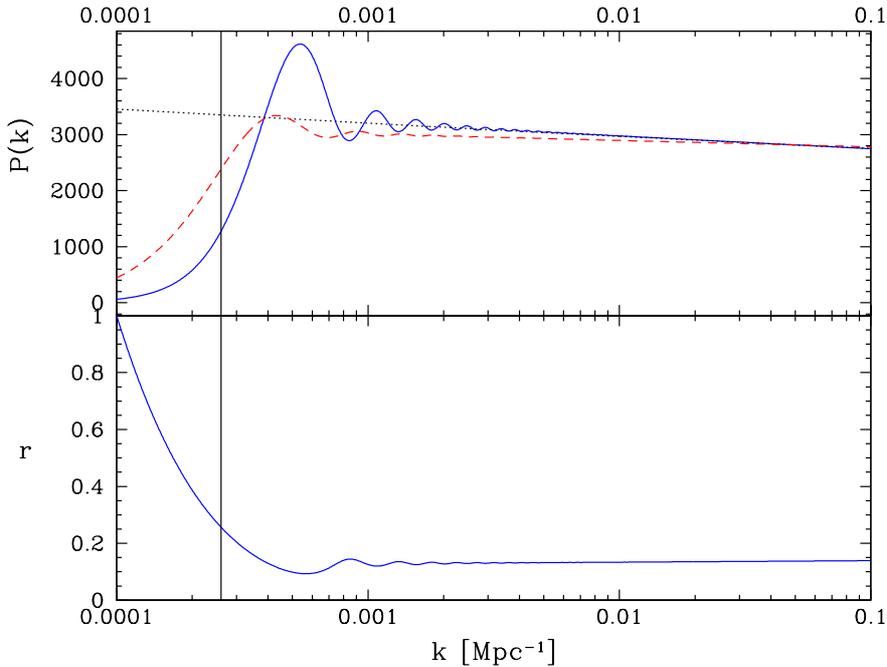}
\caption{The scalar and tensor power spectra and tensor-to-scalar
  ratio for the initially kinetic dominated trajectory with $V(\phi) = m^2\phi^2/2$. 
  The spectra
  are computed by numerically integrating the perturbation
  equations. The tensor spectrum (dashed) is rescaled for
  comparison. Both show a suppression on large scales. However the
  cut-off is sharper in the scalar modes than the tensor modes. The
  limiting power law spectral index for the scalar power spectrum is
  $n_s=0.967$ at $k=0.05$ Mpc$^{-1}$ and is shown as a dotted line in
  the top panel. The solid, vertical line indicates the scale, $k_c$,
  of the horizon at the beginning of inflation.}
\label{fig:kd}
\end{figure}

The plane wave expanded evolution equation for the variable $u$ at
each momentum wavenumber $k$ takes the form of an oscillator with
time dependent mass
\begin{equation}\label{eq:scalar}
  u_k'' + \left( k^2 - \frac{z''}{z}\right)\,u_k = 0,
\end{equation}
where $z=a\phi'/{\cal H}$. The dimensionless power spectrum of the
curvature perturbation $\Phi$ can be obtained by evolving each complex
$u_k$ mode from inside the horizon at early times to its late-time,
super-horizon solution at which point $u/a \rightarrow \dot\phi\Phi/H$
and becomes constant and
\begin{equation}
  P_S(k) = \frac{k^3}{2\pi^2} \left|\frac{u_k}{z}\right|^2 .
\end{equation} 
Each polarisation of the plane wave
expansion of the tensor perturbation ($h\equiv h_\times$, $h_+$) evolves as a
free scalar field evolving in an expanding background. The momentum
space evolution equation for the variable $v=ah$
is
\begin{equation}\label{eq:tensor}
  v_k'' + \left( k^2 - \frac{a''}{a}\right)\,v_k = 0.
\end{equation}
The dimensionless power spectrum of the tensor
modes is again obtained by the late-time, super-horizon value of the
each $v_k$ mode as 
\begin{equation}
  P_T(k) = \frac{8\,k^3}{\pi} \left|\frac{v_k}{a}\right|^2.
\end{equation}

For initial conditions we insure the inflaton is fast-rolling by 
choosing $\dot \phi_0 = 10\, m\phi_0$ in units of reduced Planck mass. 
The choice of $\phi_0$ is determined
by the potential, for $V(\phi) =  m^2\phi^2/2$ we take $\phi_0 = 18$. 
To obtain the same number of $e$-folds in the $V(\phi) = \lambda \phi^4$ 
potential we take $\phi_0 = 24$.
The initial conditions for the perturbations in a fast-roll regime can
be set by taking the limit $\dot\phi\gg\phi$ for the background with
$z''/z \approx a''/a$. In this case the solutions to
Eqs.~(\ref{eq:scalar}) and (\ref{eq:tensor})  approach \cite{Contaldi03}
\begin{equation}
f_k \left( \eta \right) = \left[\frac{\pi}{8}\frac{(1 +
2H_0\eta)}{H_0}\right]^{1/2} {\cal H}_0^{(2)} \left( k\eta +
\frac{k}{2H_0} \right)
\label{solv}
\end{equation}
where ${\cal H}_0^{(2)}$ is a Hankel's function of the second
kind. Note that we have chosen to consider only positive frequency
components of the solutions and we have normalised such that it would
recover the correct adiabatic vacuum in the small scale limit 
\begin{equation}\label{eq:vacuum}
f_k(\eta) = \frac{1}{\sqrt{2k}}e^{-ik\eta}. 
\end{equation}
This choice assumes that even modes close to or larger than the
horizon were at some point much smaller than the horizon where they
asymptote to the the Bunch Davis vacuum solution \cite{bunch}. This is
justified if for example the universe was inflating before the kinetic
dominated stage (see {\sl e.g.} \cite{Contaldi03} for a more in depth
discussion). We will discuss the alternative choice of a radiation
dominated initial stage in~\ref{sec:rd}.

\begin{figure}[t]
\centering
\includegraphics[width=9cm,angle=270]{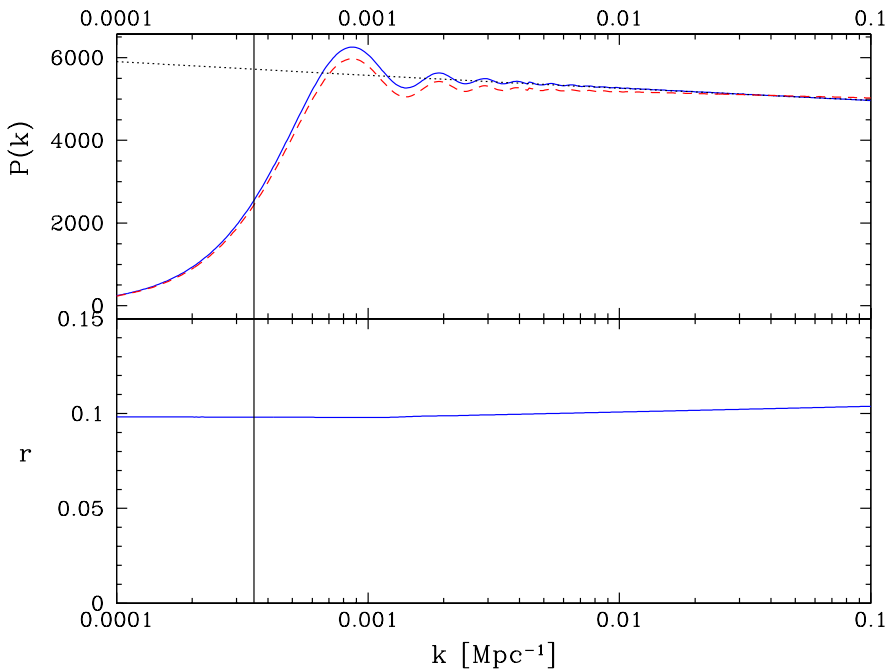}
\caption{The scalar and tensor power spectra and tensor-to-scalar
  ratio for radiation dominated initial conditions
  with $V(\phi) = m^2\phi^2/2$. The spectra
  are computed by numerically integrating the perturbation
  equations. The tensor spectrum (dashed) is rescaled for
  comparison. Both show a suppression on large scales, the form of
  which is almost indistinguishable. The
  limiting power law spectral index for the scalar power spectrum is
  $n_s=0.967$ at $k=0.05$ Mpc$^{-1}$ and is shown as a dotted line in
  the top panel. The solid, vertical line indicates the scale, $k_c$,
  of the horizon at the beginning of inflation.}
\label{fig:rd}
\end{figure}

In Fig.~\ref{fig:kd} we show the spectra that result from a numerical
integration of the $u_k$ and $v_k$ modes over a range of
wavenumbers. Both scalar and tensor spectra are plotted and the tensor
spectrum is scaled to match the scalar power at a scale well inside the
slow-roll regime. The tensor-to-scalar ratio is shown in the
bottom panel. 

The evolution of each mode is started at a fixed time in the kinetic
regime and stopped once the mode is a number of times larger than the
horizon at which point both $u_k/a$ and $v_k/a$ have reached their
constant value. Fig. ~\ref{fig:kd} reproduces the result for the
scalar spectrum found in \cite{Contaldi03} (we have made the same
choice of initial parameters and potential $V(\phi) = m^2\phi^2/2$)
with an exponential cut-off in the spectrum at large scale. The tensor
modes also show a suppression, however it is not as sharp as the
scalar suppression. Thus the ratio of the two spectra rises on the
largest scales. We have also checked that a similar result is obtained
for a $V(\phi)=\lambda\phi^4/4$ choice of inflaton potential.

As shown in \cite{Contaldi03} the cut-off result can be understood in
a simple analytical picture of an instantaneous transition from
kinetic to de Sitter regimes. The analytical solution is identical for
the scalar and tensor cases, however there are crucial differences in
their detailed evolution through the transition as shown in the
numerical case. The difference is greatest for modes that are close to
the horizon scale during the transition between regimes when $z''/z$
temporarily diverges from $a''/a$.

The qualitative result is that the super-horizon tensor modes are not
suppressed as strongly as the scalars during the initial kinetic
phase. The effect is most clearly seen in the ratio of the spectra
$r(k)=P_T(k)/P_S(k)$ which approaches $1$ on the largest scales. This
effect can also be understood within the simple slow--roll framework
which gives \cite{Liddle}
\begin{equation}
  \frac{P_T}{P_S}\sim \dot\phi^2,
\end{equation}
although the approximation is only valid for small values of
$\dot\phi$. The scalar (curvature) perturbation is always suppressed by
the extra $\dot\phi \sim 1/\dot H$ term with respect to the tensor
modes and this effect becomes most pronounced if inflation started
from a KD stage.

\subsection{Radiation dominated initial conditions}\label{sec:rd}

We now compare this result to the case where a radiation dominated
stage preceded the stage of inflation. We assume that the inflaton is
already in the slow roll regime when the inflaton energy density
begins to dominate. We carry out a numerical mode-by-mode integration
as in the previous section. The initial conditions for the modes are
modified in this case \cite{Kinney06,linderd} by using the adiabatic
vacuum condition identical to Eq.~\ref{eq:vacuum}.

Fig.~\ref{fig:rd} shows the scalar and tensor spectra obtained for the
radiation dominated initial conditions. Both spectra now show similar
cutoffs and the ratio remains roughly constant. It is important to
note that this difference is simply due to the different initial
conditions chosen for the inflaton's kinetic term rather than being
due to the normalisation of the perturbations. In fact the cut-off
picture is quite insensitive to the exact choice of normalisation
which inevitably involves some assumptions about the state of the
perturbations before the start of inflation.

\begin{figure}[t]
\centering
\includegraphics[width=12cm]{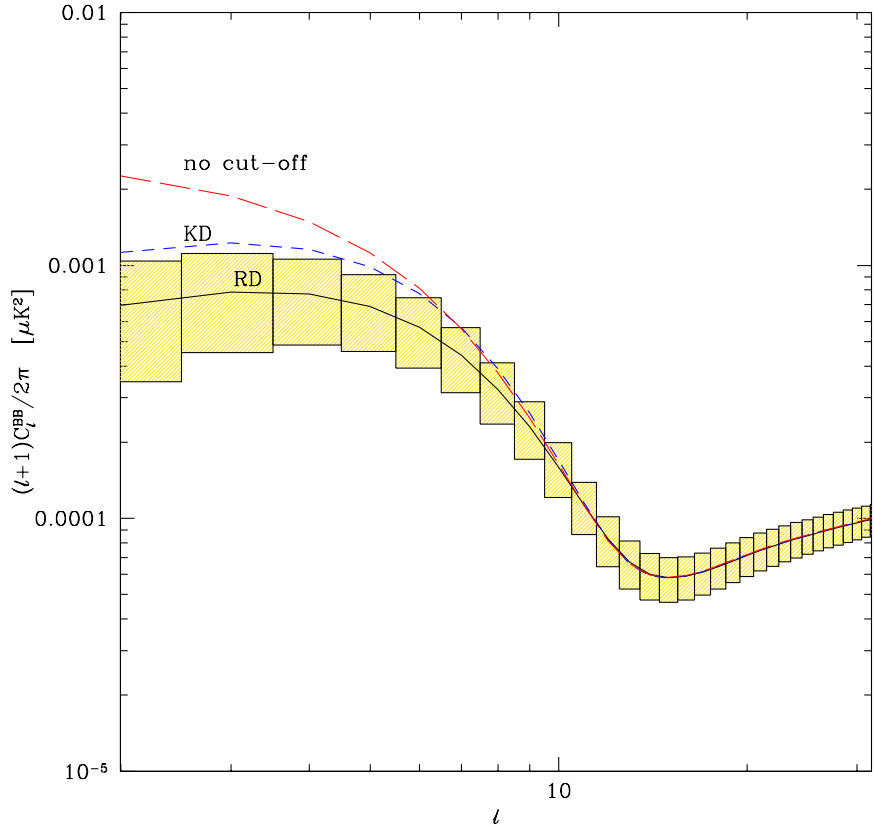}
\caption{The ratio ${\cal R}_\ell=C_\ell^{BB}/C_\ell^{EE}$ for a
  standard, slow-roll (SR), power law spectra case (black, solid) and
  for the model with a kinetic cut-off (blue, dashed), top
  panel. ${\cal R}_\ell$ is enhanced at low multipoles due to the
  weaker suppression of tensor modes during the transition from the
  kinetic to inflating regime. The bottom panel shows the ratio of the
  two curves. We also plot naive, sample--variance error boxes for
  bins of width $\Delta \ell=3$. We have assumed an optimal
  polarisation experiment observing 80\% of the sky. The cut-off scale
  was placed at $k_c=5.3\times 10^{-4}$ Mpc$^{-1}$ for this example.}
\label{fig:rdl}
\end{figure}

\section{Observability}\label{sec:observe}

Although the difference between kinetic dominated and non-kinetic
dominated initial conditions is quite pronounced in the amplitude of
the tensors it will only affect the largest scales observed in the CMB
today. Thus the accuracy with which the effect can be observed is
limited by cosmic variance. 

The next generation of CMB polarisation satellites will aim to survey
the whole sky with sufficient resolution and sensitivity to be sample
variance limited on the largest angular scales. It is therefore
timely to ask whether future missions will be able to detect a cutoff
and distinguish between the KD and RD scenarios. To produce a figure of
merit we have assumed a future experiment which returns a sample
variance limited $C_\ell^{BB}$ power spectrum at $\ell < 100$. We assume a
coverage of 80\% of the sky, as this is probably the best one can hope
to use for polarisation measurements.

The resulting sample variance errors at each multipole in the B-mode
power spectrum are shown in Fig.~\ref{fig:rdl}. The three curves shown
are the B-mode spectra for a model with no cut-off and both KD and RD
models discussed above. All three models have identical scalar
spectral indexes ($n_s=0.967$) on small scales and are normalised such
that the total power at the first acoustic peak matches the WMAP
3-year results \cite{Hinshaw06}. The model angular power spectra with
the cutoffs are obtained using a version of the {\tt CAMB} code
\cite{camb} where the modified power spectra are used instead of a
simple power law for the initial conditions. No other modifications is
required in the line of sight codes since radiation transfer is
unaffected in these models.

As an indication of the level at which the two scenarios can be
distinguished we calculate the $\chi^2$ value for the difference
between the RD and KD curves and the RD and no cut-off curves. We
obtain $\chi^2= 5.03$ and $\chi^2= 8.17$ respectively.

The dependence seen here is degenerate with the effect of reionisation
on the polarisation spectra. Our models assume an optical depth
$\tau=0.9$, however the value of $\tau$ will be fixed to high
accuracies by the E-mode and TE cross-correlation spectra for such an
experiment and it is reasonable to expect that the degeneracy will be
broken if one allows for a cut-off in the primordial
spectrum. Alternatively if a cut-off is not taken into account in
parameter fits the effect would show up as a mismatch between the
value of $\tau$ required by the E-mode spectrum and that required by
the B-modes.

The quest for such a detection is certainly an ambitious one. A
detailed statistical forecast using the full parameter space Fisher
matrix would be required to establish the true signal-to-noise of the
effect. There are, of course, phenomenal obstacles to overcome before
experiments measure the B-mode $C_\ell$ spectrum to such precision on
the largest scales.  These include a significant increase in
instrumental sensitivity, but most of all, the level of polarised
foreground contamination is yet to be determined. Indeed it is still
to be established whether this naive, sample limited result can ever
be reached. However the first generation of sub-orbital experiments
targeting the B-mode $C_\ell$ are getting underway. These
observations, at the very least, will determine to what accuracy this
region of the spectrum can be measured. These include balloon-borne
experiments such as {\tt SPIDER} \cite{spider},
now funded, which will cover half the sky and will start to probe the
low-$\ell$ region of the B-mode spectrum. Ultimately though the region
will be probed most accurately by the next generation of CMB
satellites (see e.g. \cite{cmbpol}).

\section{Discussion}\label{sec:disc}

Given the rapidly advancing observational front of CMB polarisation
experiments it is timely to examine the detailed predictions of
cut-off models and their observational consequences. We have shown
here how the tensor-to-scalar ratio of primordial perturbations
depends on the nature of the pre-inflationary epoch in short inflation
models which give rise to a cut-off on large scales. The different
scenarios may be distinguishable once sample variance limited
observations of the CMB polarisation B-modes are made. The
detectability of the effect also depends on the amplitude of the
B-mode spectrum which can be much smaller than the ratio $r=0.1$
considered here. In particular if multiple fields are present during
inflation then the tensor amplitude may be too low compared with
foreground levels to be observable.

If the cut-off is confirmed to be present in the B-mode spectrum too
however it will pose some fundamental questions about the initial
conditions for inflation and the question of apparent fine tuning will
become even more pressing.  At the very least it would pose
significant questions as to the appropriate prior to adopt when
constraining the full space of inflationary solutions
\cite{trajectories}.

An interesting consequence of the canonical measure picture is that if
we accepted {\sl a priori} to live on one of the rare solutions that
achieved say $N=60$ $e$-foldings then we should expect inflation to
have lasted just {\sl that} long. Thus trajectories with KD initial
conditions examined in detail here are precisely the kind which
dominate in the canonical measure picture.  Similar trajectories seem
to be suggested by simple probabilistic analysis of inflation models
on the string theory landscape. Both scenarios however rely on
anthropic considerations to justify the selection of extremely rare
trajectories. 

It is important to note that we have kept the assumptions of
homogeneous and isotropic background in examining the signature of
short inflation. These are not the most general initial
conditions. Indeed inflation may have started in an inhomogeneous and
anisotropic background and such initial conditions may lead to even
stronger effects in the observable modes or potentially rule out
inflation that is ``just short enough''
\cite{emir,donoghue,perreira}. However, in both cases and in the not
so distant future, the lowest CMB multipoles may yet reveal important
clues to the nature of inflation.

\acknowledgements

We are grateful to Dick Bond, Lev Kofman and Marco Peloso for
enlightening discussions. This work was supported by a STFC
studentship.


\begin{thebibliography}{99}
\expandafter\ifx\csname natexlab\endcsname\relax\def\natexlab#1{#1}\fi
\expandafter\ifx\csname bibnamefont\endcsname\relax
  \def\bibnamefont#1{#1}\fi
\expandafter\ifx\csname bibfnamefont\endcsname\relax
  \def\bibfnamefont#1{#1}\fi
\expandafter\ifx\csname citenamefont\endcsname\relax
  \def\citenamefont#1{#1}\fi
\expandafter\ifx\csname url\endcsname\relax
  \def\url#1{\texttt{#1}}\fi
\expandafter\ifx\csname urlprefix\endcsname\relax\def\urlprefix{URL }\fi
\providecommand{\bibinfo}[2]{#2}
\providecommand{\eprint}[2][]{\url{#2}}

 \bibitem[Hinshaw et al.(2007)]{Hinshaw06} Hinshaw, G., et al.\ 
2007, \apjs, 170, 288 

 \bibitem[Efstathiou(2004)]{efstathiou} Efstathiou, G.\ 2004, 
\mnras, 348, 885 

 \bibitem[de Oliveira-Costa et al.(2004)]{costa} de 
Oliveira-Costa, A., Tegmark, M., Zaldarriaga, M., \& Hamilton, A.\ 2004, 
\prd, 69, 063516 

 \bibitem[Magueijo \& Sorkin(2006)]{magueijo} Magueijo, J., \& 
Sorkin, R.~D.\ 2006, ArXiv Astrophysics e-prints, arXiv:astro-ph/0604410 

s
 \bibitem[Bielewicz et al.(2004)]{bielewicz} Bielewicz, P., 
G{\'o}rski, K.~M., \& Banday, A.~J.\ 2004, \mnras, 355, 1283 

 \bibitem[Bridle et al.(2003)]{bridle} Bridle, S.~L., Lewis, 
A.~M., Weller, J., \& Efstathiou, G.\ 2003, \mnras, 342, L72 

 \bibitem[Schwarz et al.(2004)]{schwarz} Schwarz, D.~J., 
Starkman, G.~D., Huterer, D., \& Copi, C.~J.\ 2004, Physical Review 
Letters, 93, 221301 

 \bibitem[Slosar et al.(2004)]{slosar} Slosar, A., Seljak, U., 
\& Makarov, A.\ 2004, \prd, 69, 123003 

 \bibitem[Prunet et al.(2005)]{prunet} Prunet, S., Uzan, J.-P., 
Bernardeau, F., \& Brunier, T.\ 2005, \prd, 71, 083508 

 \bibitem[Contaldi et al.(2003)]{Contaldi03} Contaldi, C.~R., 
Peloso, M., Kofman, L., \& Linde, A.\ 2003, Journal of Cosmology and 
Astro-Particle Physics, 7, 2 

 \bibitem[Cline et al.(2003)]{Cline03} Cline, J.~M., Crotty, P., 
\& Lesgourgues, J.\ 2003, Journal of Cosmology and Astro-Particle Physics, 
9, 10 

 \bibitem[Kaloper(2004)]{Kaloper04} Kaloper, N.\ 2004, Physics 
Letters B, 583, 1 

 \bibitem[Piao et al.(2004)]{Piao04} Piao, Y.-S., Feng, B., \& 
Zhang, X.\ 2004, \prd, 69, 103520 

 \bibitem[Powell \& Kinney(2006)]{Kinney06} Powell, B.~A., \& 
Kinney, W.~H.\ 2006, ArXiv Astrophysics e-prints, arXiv:astro-ph/0612006 

 \bibitem[Linde(1983)]{chaotic} Linde, A.~D.\ 1983, Physics 
Letters B, 129, 177 

 \bibitem[Kofman et al.(2002)]{Kofman02} Kofman, L., Linde, A., 
\& Mukhanov, V.\ 2002, Journal of High Energy Physics, 10, 57 

 \bibitem[Gibbons et al.(1987)]{Gibbons86} Gibbons, G.~W., 
Hawking, S.~W., \& Stewart, J.~M.\ 1987, Nuclear Physics B, 281, 736 

 \bibitem[Hawking \& Page(1988)]{Hawking88} Hawking, S.~W., \& 
Page, D.~N.\ 1988, Nuclear Physics B, 298, 789 

 \bibitem[Gibbons \& Turok(2006)]{Gibbons06} Gibbons, G.~W., \& 
Turok, N.\ 2006, ArXiv High Energy Physics - Theory e-prints, 
arXiv:hep-th/0609095 

 \bibitem[Freivogel et al.(2006)]{freivogel} Freivogel, B., 
Kleban, M., Rodr{\'{\i}}guez Mart{\'{\i}}nez, M., \& Susskind, L.\ 2006, 
Journal of High Energy Physics, 3, 39 

 \bibitem[Coleman \& de Luccia(1980)]{coleman} Coleman, S., \& 
de Luccia, F.\ 1980, \prd, 21, 3305 

 \bibitem[Douglas(2003)]{douglas} Douglas, M.~R.\ 2003, Journal 
of High Energy Physics, 5, 46 

 \bibitem[Gumrukcuoglu et al.(2007)]{emir} Gumrukcuoglu, 
A.~E., Contaldi, C.~R., \& Peloso, M.\ 2007, ArXiv e-prints, 707, 
arXiv:0707.4179

 \bibitem[Mukhanov et al.(1992)]{Mukhanov} Mukhanov, V.~F., 
Feldman, H.~A., \& Brandenberger, R.~H.\ 1992, \physrep, 215, 203 

 \bibitem[Bunch and Davies (1978)]{bunch} Bunch, T.~S., \& Davies, P.~C.~W,
1978, Proceedings of the Royal Society of London, Series A, 10, 117 

 \bibitem[Liddle \& Lyth(2000)]{Liddle} Liddle, A.~R., \& Lyth, 
D.~H.\ 2000, Cosmological Inflation and Large-Scale Structure, by Andrew 
R.~Liddle and David H.~Lyth, pp.~414.~ISBN 052166022X.~Cambridge, UK: 
Cambridge University Press, April 2000.,

 \bibitem[Linde(1982)]{linderd} Linde, A.~D.\ 1982, Physics 
Letters B, 116, 335 

 \bibitem[Lewis et al.(2000)]{camb} Lewis, A., Challinor, A., 
\& Lasenby, A.\ 2000, \apj, 538, 473 

 \bibitem[Montroy et al.(2006)]{spider} Montroy, T.~E., et al.\ 
2006, \procspie, 6267

 \bibitem[Bock et al.(2006)]{cmbpol} Bock, J., Hinshaw, G.~F., 
 \& Timbie, P.~T.\ 2006, Bulletin of the American Astronomical Society, 38, 
963 

 \bibitem[{\citenamefont{{Bond},{Contaldi},{Kofman} and {Vaudrevange}}
    in preparation, 2007}]{trajectories}
 \bibinfo{author}{\bibfnamefont{J.}~\bibfnamefont{R.}~\bibnamefont{{Bond}}},
    \bibfnamefont{C.}~\bibfnamefont{R.}~\bibnamefont{{Contaldi}},
    \bibfnamefont{L.}~\bibnamefont{{Kofman}} and \bibfnamefont{P.}~\bibnamefont{{Vaudrevange}},
    in preparation (2007).
\bibitem[Donoghue et al.(2007)]{donoghue} Donoghue, J.~F., 
Dutta, K., \& Ross, A.\ 2007, ArXiv Astrophysics e-prints, 
arXiv:astro-ph/0703455 
\bibitem[Pereira et al.(2007)]{perreira} Pereira, T.~S., Pitrou, 
C., \& Uzan, J.-P.\ 2007, Journal of Cosmology and Astro-Particle Physics, 
9, 6 


\end{thebibliography}
\end{document}